\documentclass{ws-mpla}

\usepackage[super]{cite}
\usepackage{xcolor}
\usepackage[verbose,hypertexnames=false]{hyperref}
\hypersetup{colorlinks=false,allbordercolors=blue,pdfborderstyle={/S/U/W 1}}
\usepackage{graphicx}
\usepackage{amsmath}
\usepackage{amssymb}
\graphicspath{{./images/}}

\begin{document}

\markboth{F. Shipilov et al.}{ML-based Fast Simulation of FARICH Responses}

\catchline{}{}{}{}{}

\title{ML-based Fast Simulation of FARICH Responses}

\author{F. Shipilov\footnote{foma@shipilov.ru}}

\address{HSE University, Moscow, Russia}

\author{A. Barnyakov}

\address{Budker Institute of Nuclear Physics SB RAS, Novosibirsk, Russia\\
Novosibirsk State Technical University, Novosibirsk, Russia}

\author{A. Ivanov}

\address{Joint Institute for Nuclear Research, Dubna, Russia}

\author{F. Ratnikov}

\address{HSE University, Moscow, Russia}

\maketitle

\begin{history}
\received{(Day Month Year)}
\revised{(Day Month Year)}
\accepted{(Day Month Year)}
\published{(Day Month Year)}
\end{history}

\begin{abstract}
A fast simulation of the detector response is a vital task in high-energy physics (HEP).
Traditional Monte-Carlo methods form the backbone of modern particle physics simulation
software but are computationally expensive. We present a machine-learning-based approach
to fast simulation of the Focusing Aerogel Ring Imaging Cherenkov (FARICH) detector
response. Given a particle track and momentum, the goal is to generate realistic samples
of photon hits on the detector matrix. We propose a conditional Generative Adversarial
Network (cGAN) with a lightweight convolutional architecture that
reproduces the projected detector response conditioned on particle parameters. We compare
the cGAN against a linear statistical baseline using metrics
applied to probability maps and to the reconstructed velocity distributions. The cGAN provides a high inference throughput and a good agreement with Geant4 reference. Further work is needed to ensure complete reconstructed velocity reproduction.
\end{abstract}


\ccode{PACS Nos.: 29.40.Ka, 29.85.Fj}

\section{Introduction}
\label{sec:intro}

The Spin Physics Detector (SPD) is a universal detector at the
Nuclotron-based Ion Collider fAcility (NICA), proposed to study Drell--Yan processes,
J/$\Psi$ production processes, elastic reactions, spin effects in one- and two-hadron
production, polarization effects in heavy-ion collisions, and more.\cite{savin2015spin}
The SPD is a medium-energy experiment with luminosity up to $10^{32}$~cm$^{-2}$s$^{-1}$
and a free-flowing (triggerless) running mode, which requires novel approaches to data
acquisition.\cite{abazov2021conceptual}

Reliable particle identification (PID) is a crucial component of modern HEP
experiments. An aerogel counter in the SPD end-cap region has been proposed to
provide $\pi$/K-separation below 5~GeV/c using a Focusing Aerogel Ring Imaging
Cherenkov (FARICH) detector.\cite{thespd2023technical,korzenev2023spin}
FARICH uses a multilayer aerogel stack with increasing refractive index for proximity
focusing, thus removing the need for focusing mirrors while preserving a good
Cherenkov angle resolution.\cite{KINDO2020162252}

Machine learning (ML) techniques have been shown to be state-of-the-art across a
broad range of HEP problems,\cite{albertsson2018machine,Boehnlein_2022} including
calibration and reconstruction of Cherenkov detectors.\cite{Fanelli_2020}
Existing RICH PID methods exploit various statistics of the Cherenkov angle
distribution; e.g. the LHCb\cite{POWELL2011260} and AMS\cite{BARAO2003310}
experiments discriminate mass hypotheses via maximum-likelihood estimation, while
Belle~II uses a Gaussian approximation.\cite{KINDO2020162252}
Neural networks have also been applied to RICH PID in LHCb\cite{blago2023deep}
and in end-to-end reconstruction frameworks for DIRC
detectors.\cite{fanelli2020deeprich,fanelli2025deep}

A fast simulation of the detector response is growingly important for HEP experiments. Computer
simulations provide a detailed theoretical reference against which models of both
known and ``new'' physics can be tested. Traditional Monte-Carlo methods at the
backbone of modern simulation software (e.g.~Geant4\cite{agostinelli2003geant4,%
allison2006geant4}) are highly accurate but computationally expensive. Deep
generative models can leverage inherent symmetries and effectively factorize the
state space of the detector response, providing a significant speed-up while
preserving important physics relationships.\cite{ratnikov2020generative,%
ratnikov2021fast,chekalina2019generative,derkach2020cherenkov,%
srebre2020generation,sergeev2021fast,rogachev2023gan}
Recently, generative models have been applied specifically to fast simulation of Detection of Internally Reflected Cherenkov Light (DIRC) detectors at LHCb and future Electron-Ion Collider (EIC), demonstrating
that GANs can reproduce Cherenkov ring statistics with high
fidelity.\cite{derkach2020cherenkov,giroux2025generative}

In this work, we develop an ML-based fast simulation for the FARICH detector.
Given particle track parameters and momentum, the aim is to generate realistic
samples of Cherenkov photon hits on the detector matrix, bypassing the slow
photon-by-photon Monte-Carlo process. We employ a conditional GAN
(cGAN)\cite{mirza2014conditional} with a lightweight convolutional architecture
and compare it against a linear statistical baseline.

\section{Overview of the data}
\label{sec:data}

RICH detectors provide PID by measuring the trajectories of Cherenkov photons. An ultrarelativistic particle emits
Cherenkov photons when it traverses the transparent radiator. The photons are emitted within a cone whose spread angle $\theta_c$ (Cherenkov angle) is a function of the particle velocity
$\beta = v/c$. Together with momentum
$p$ from the tracking system, measuring this angle bounds the mass of the particle and enables particle identification. In FARICH, a multilayer aerogel with graded refractive index is used as a radiator, focusing Cherenkov photons onto a flat photon detector matrix. The photosensitive matrix is a flat grid of silicon photomultipliers (SiPMs).
Cherenkov photons from the aerogel radiator strike the matrix at positions that
depend on the Cherenkov emission angle and the particle track geometry,
forming an approximately elliptical pattern on the detector surface.

Fast simulation aims to speed up the traditional simulation by fitting a surrogate model over a dataset. In this work, we solve a conditional simulation problem, which is similar to the initial one solved by Monte-Carlo. For conditional simulation, each event is characterized by:
\begin{itemlist}
  \item \emph{Input parameters} (conditions): particle entry coordinates $(x_p, y_p, z_p)$,
    polar angle $\theta_p$ and azimuth $\phi_p$ of the particle direction, velocity $\beta$ and momentum $p$;
  \item \emph{Output sample} (detector response): coordinates of triggered pixels
    $(x_c, y_c, z_c)$ on the photosensitive matrix.
\end{itemlist}

In this work, we train our models on a dataset that was produced with Geant4 simulation
toolkit.\cite{agostinelli2003geant4,allison2006geant4} The dataset consisted of 27\,000\,000 secondary particle traversal events through FARICH end-caps. Motivated by the $\pi$/K separation objective of FARICH, we focus on these two particle species. Their relative sample sizes in the data were $95\%/5\%$ for $\pi$/K respectively.

\section{Methods}
\label{sec:methods}

\subsection{Data preprocessing}
\label{subsec:preprocess}

Cherenkov photons experience refraction at the aerogel--air
interface, which causes the measured hit positions to deviate from the ideal
Cherenkov cone. A key preprocessing step for our models is the correction of these optical effects. We utilize a previously developed numerical method for refraction correction based on the fixed-point iteration.\cite{shipilov2024reconstruction} Crucially, this enables projecting photon hits into a $64\times64$ grid representing the
Cherenkov ring in polar coordinates, yielding a compact 2D image suitable for
convolutional processing. This representation factors out geometrical effects, including rotation of the ring, allowing the surrogate model to focus on the physical structure of the sample. Simulated FARICH responses in this format can then be projected back to the original photon detector matrix for subsequent analysis.

Additionally, we optimize the parameters used in the refraction correction formula, namely the value of the refractive index and the position of the photon cone origin inside the aerogel radiator. This removes the systematic bias that was present earlier and further improves the resolution of the method.

\subsection{Linear simulation baseline}
\label{subsec:linear}

The linear baseline assumes the independence of the Cherenkov angle
$\theta_c$ from the azimuth $\phi_c$ (after refraction correction). While this is a flawed assumption, it is necessary for the proper functioning of a linear model. A ridge linear regression is fitted to
predict the mean $\mu$ and standard deviation $\sigma$ of the $\theta_c$
distribution as a function of the true particle parameters
($\beta^{\rm true}$, $\theta_p^{\rm true}$,
$p^{\rm true}$) and derivative features, separately for pions and kaons to account for the difference in their momentum distributions. At inference time, $\theta_c$ is
sampled from $\mathcal{N}(\mu, \sigma^2)$ and $\phi_c$ is drawn uniformly
from $[0, 2\pi)$. The resulting hit coordinates are then quantized to the
detector pixel grid.

\subsection{Conditional GAN}
\label{subsec:cgan}

Generative Adversarial Networks (GANs)\cite{goodfellow2014generative} involve two neural networks trained simultaneously in an adversarial setup. The \emph{generator} maps samples from
a simple prior distribution (e.g.\ a multivariate Gaussian) to outputs that are trained to
resemble real data. The \emph{discriminator} receives both real data and generator
outputs as input and is trained to tell them apart. Because the generator's loss is
constructed as the negative of the discriminator's, the two networks compete: a stronger
discriminator forces the generator to produce more realistic outputs, and vice versa.
This flexible, unsupervised objective makes GANs particularly well suited to reproducing
complex multi-modal distributions, such as those arising from detector physics, without
explicit assumptions about the form of the distribution. We adopt a conditional GAN
(cGAN)\cite{mirza2014conditional} architecture with the generator conditioned on
particle parameters ($\beta^{\rm true}$, $\theta_p^{\rm true}$, $p^{\rm true}$) to
produce a response for the specified particle kinematics.

The original GAN formulation minimizes Jensen--Shannon divergence between the real
and generated distributions.\cite{goodfellow2014generative} However, for distributions with disjoint or nearly disjoint supports, the Jensen-Shannon divergence can saturate, leading to uninformative or vanishing gradients. Alternative distance measures, in particular, the Wasserstein and Cramér distances, provide a smooth divergence even for disjoint distributions,
prevent mode collapse (where the generator covers only part of the data distribution),
and eliminate vanishing gradients.\cite{bellemare2017cramer} The Cram\'{e}r metric has
the additional benefit of yielding unbiased gradients, unlike the Wasserstein
metric.\cite{bellemare2017cramer} After careful experimental consideration of several GAN metrics, we chose the Cram\'{e}r GAN
loss\cite{bellemare2017cramer} as the adversarial component of our training objective, mainly due its superior numerical stability and sensitivity to training hyperparameter values.

As a real-valued estimator, CNN can experience difficulties reproducing the exact binary state of a photon-detector pixel. The outputs can generally be binarized using some threshold value, however, in the case of FARICH responses this biases hit coordinates. Instead, we split the generation procedure into two steps. First, the generator produces a
$64\times64$ \emph{probability map}: for each pixel of the projected ring image the
network outputs a scalar in $[0, 1]$ representing the probability of a photon hit at
that pixel. The probability map parameterizes a
Bernoulli distribution over the pixel grid and is estimated using a pixel-wise binary cross-entropy
\begin{table}[b]
    \tbl{Fast simulation results. The uncertainties are obtained from averaging model runs with different random seeds. Lower is better.\label{tab:results}}
    {\tabcolsep9.5pt\begin{tabular}{@{}lcccc@{}}
            \toprule
            Method & $\pi$ ${\rm MSE}/10^{-8}$ $\downarrow$ & K ${\rm MSE}/10^{-8}$ $\downarrow$ & Total ${\rm MSE}/10^{-8}$ $\downarrow$ & $\delta_{\rm RECO}$ $\downarrow$ \\
            \colrule
            Linear baseline & $4.1 \pm 0.1$ & $11.3 \pm 0.1$ & $4.6 \pm 0.1$ & $0.05 \pm 0.01$ \\
            \textbf{cGAN}   & $\mathbf{1.5 \pm 0.1}$ & $\mathbf{3.4 \pm 0.1}$ & $\mathbf{1.6 \pm 0.1}$ & $\mathbf{0.02 \pm 0.01}$ \\
            \botrule
    \end{tabular}}
\end{table}
\begin{figure}[b]
    \centerline{%
        \includegraphics[height=0.32\textwidth]{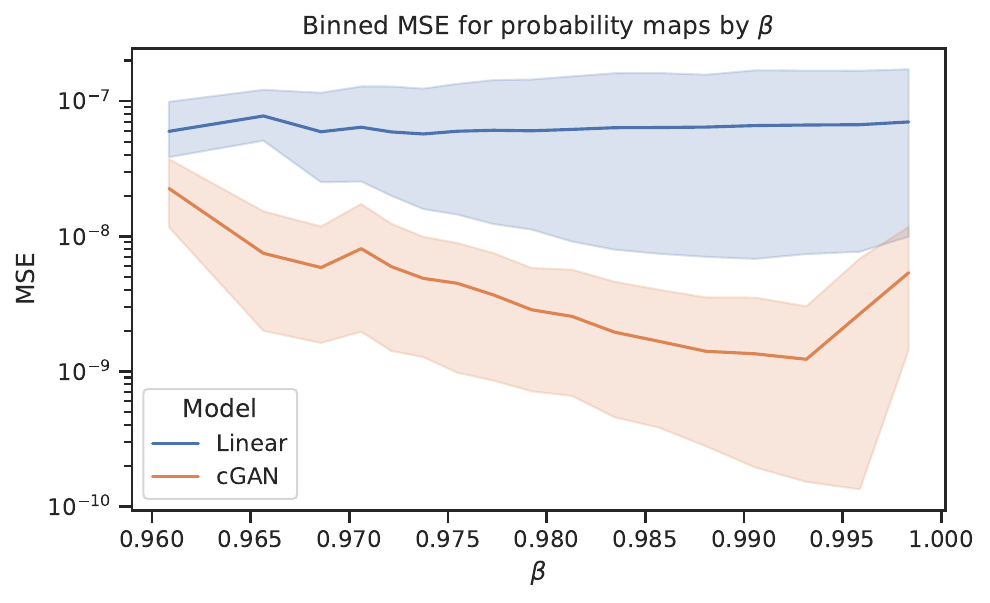}%
        \includegraphics[height=0.32\textwidth]{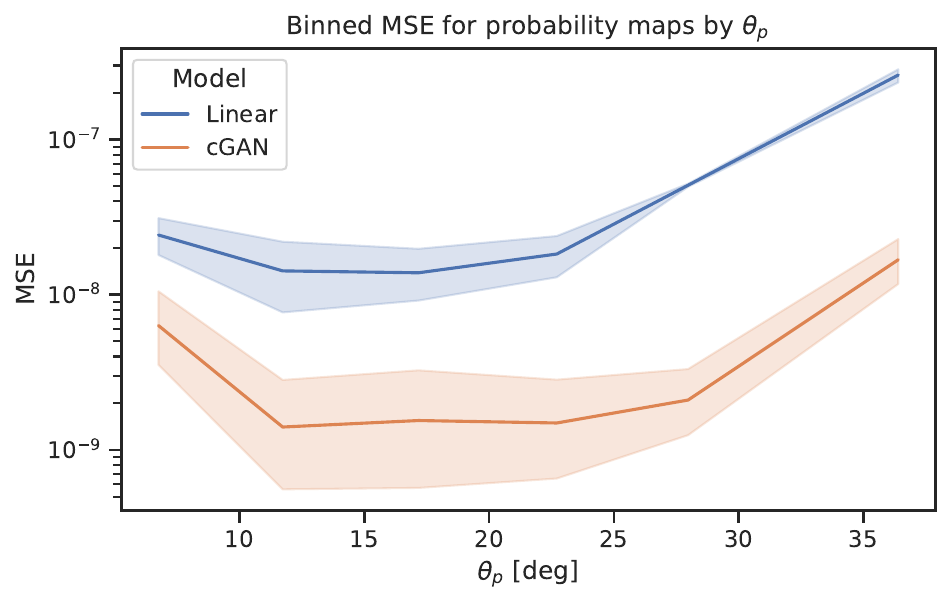}}
    \caption{MSE as a function of $\beta$ (left) and $\theta_p$ (right). The uncertainty in the plot denotes a 95\% confidence interval. It originates from the averaging across multiple events with either the same $\beta$ or $\theta_p$ parameter.}
    \label{fig:binned}
    \alttext{Plots for MSE metric.}
\end{figure}
loss as a supervised component. In the second
step, a concrete detector response is obtained at inference time by independently
sampling each pixel from its corresponding Bernoulli distribution, yielding a sparse
binary hit pattern that faithfully reflects the stochastic nature of photon detection.

The generator is a lightweight convolutional neural network (CNN) with five
upsampling stages, each consisting of two layers, totaling
810k parameters. Training employs a compound loss
combining the supervised binary cross-entropy component described above and the
adversarial Cram\'{e}r GAN loss. The model is trained for 26\,214\,400 samples with
Adam optimizer, a cosine-annealing learning rate schedule
(${\rm lr} = 10^{-4}$ for both the generator and discriminator), and a batch size of~128. A validation set of 12\,266 samples is used for overfitting control. A final testing set of 669\,402 samples is independent from both training and validation.

The relative sample sizes of $\pi$/K result in a significant class imbalance in the training set. We evaluated several mitigation strategies, including rebalancing the training set by particle type and $\beta,\theta_p$ values, both separately and at the same time, as well as loss function weighting, but did not see any improvement over the default, no intervention approach. On the contrary, the quality suffered slightly. We hypothesize that skewing the correct Geant4 distributions of $\beta$, $\theta_p$ and momentum pushes the neural network samples away from the real data.

\section{Results}
\label{sec:results}

We evaluate both fast-simulation methods using two downstream metrics. The first metric, MSE, is a mean square error of probability maps. MSE measures the closeness of the maps between the simulated and reference datasets.

\begin{figure}[t]
    \centerline{%
        \includegraphics[height=0.31\textwidth]{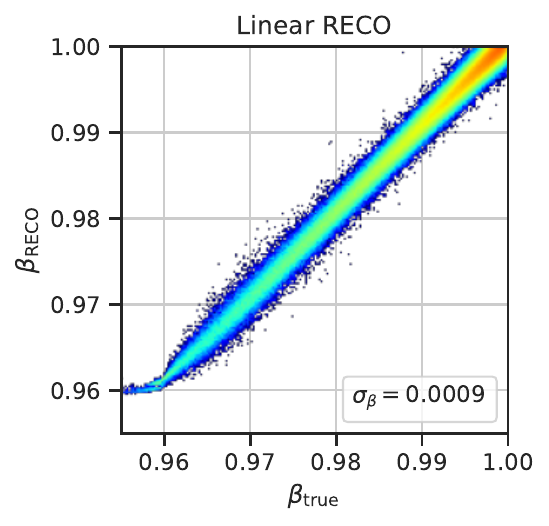}%
        \includegraphics[height=0.31\textwidth]{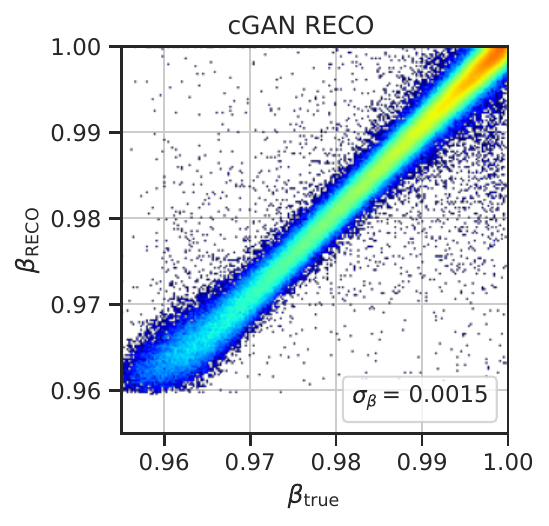}%
        \includegraphics[height=0.31\textwidth]{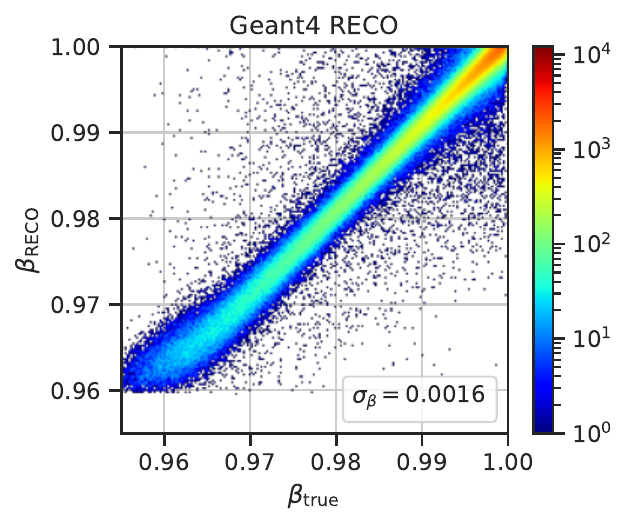}}
    \caption{2D correlation plots of reconstructed velocity from $\delta_{\rm RECO}$ computation.
        Left to right: linear baseline, cGAN, Geant4 reference.}
    \label{fig:corr_pdm}
    \alttext{Correlation plots for reconstructed velocity.}
\end{figure}

\begin{figure}[t]
{\centering
    \includegraphics[width=\textwidth]{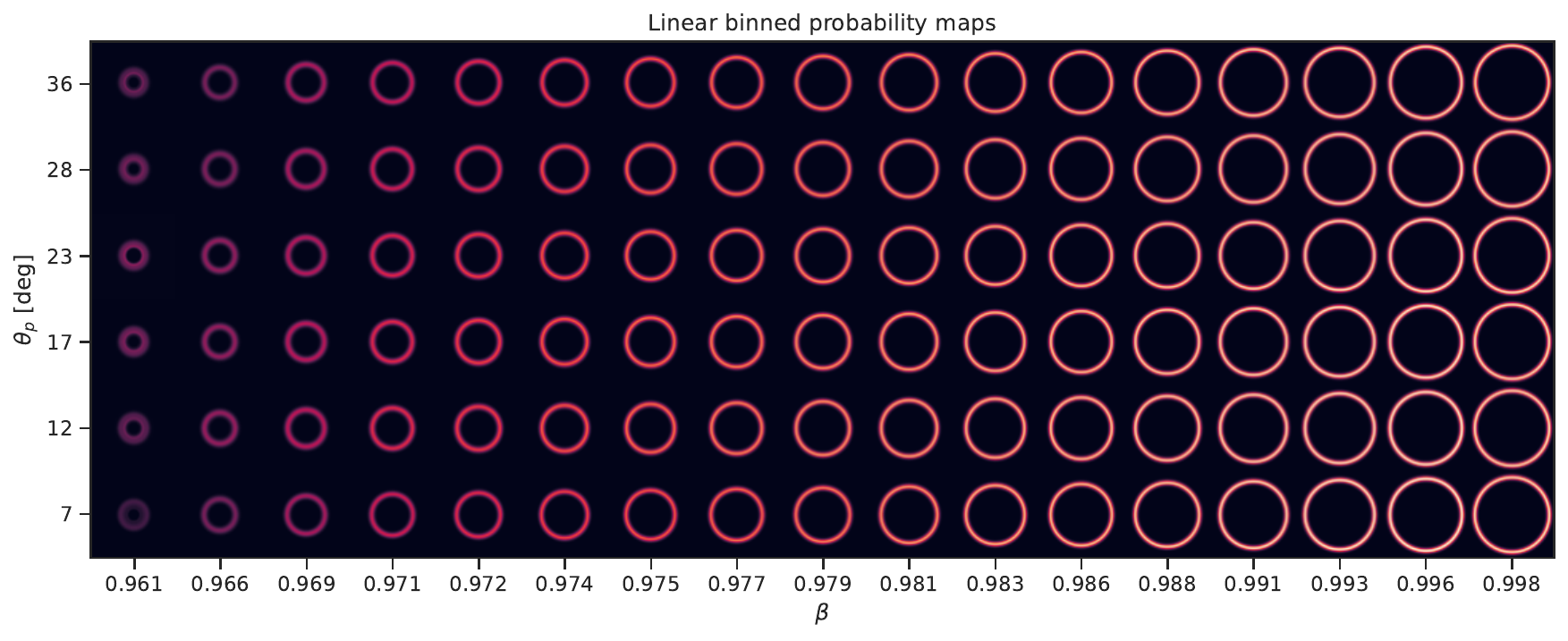}\\
    \includegraphics[width=\textwidth]{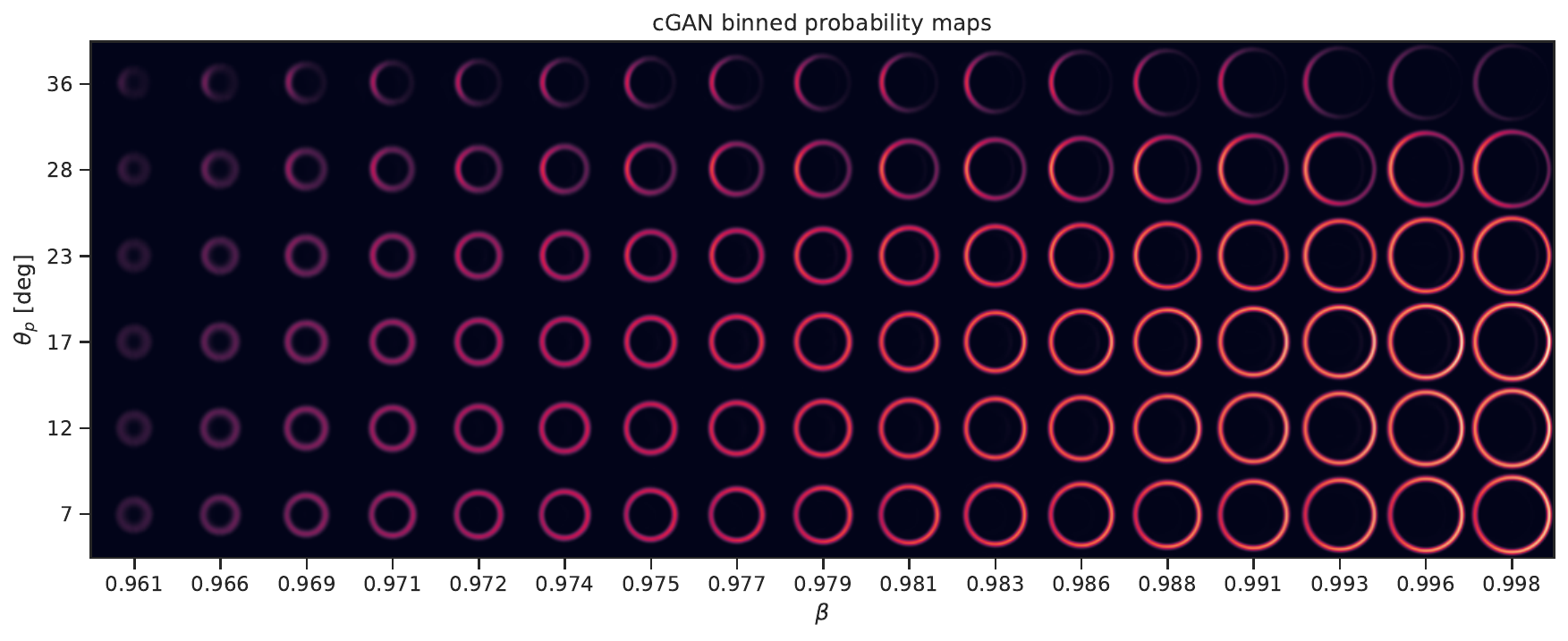}\\
    \includegraphics[width=\textwidth]{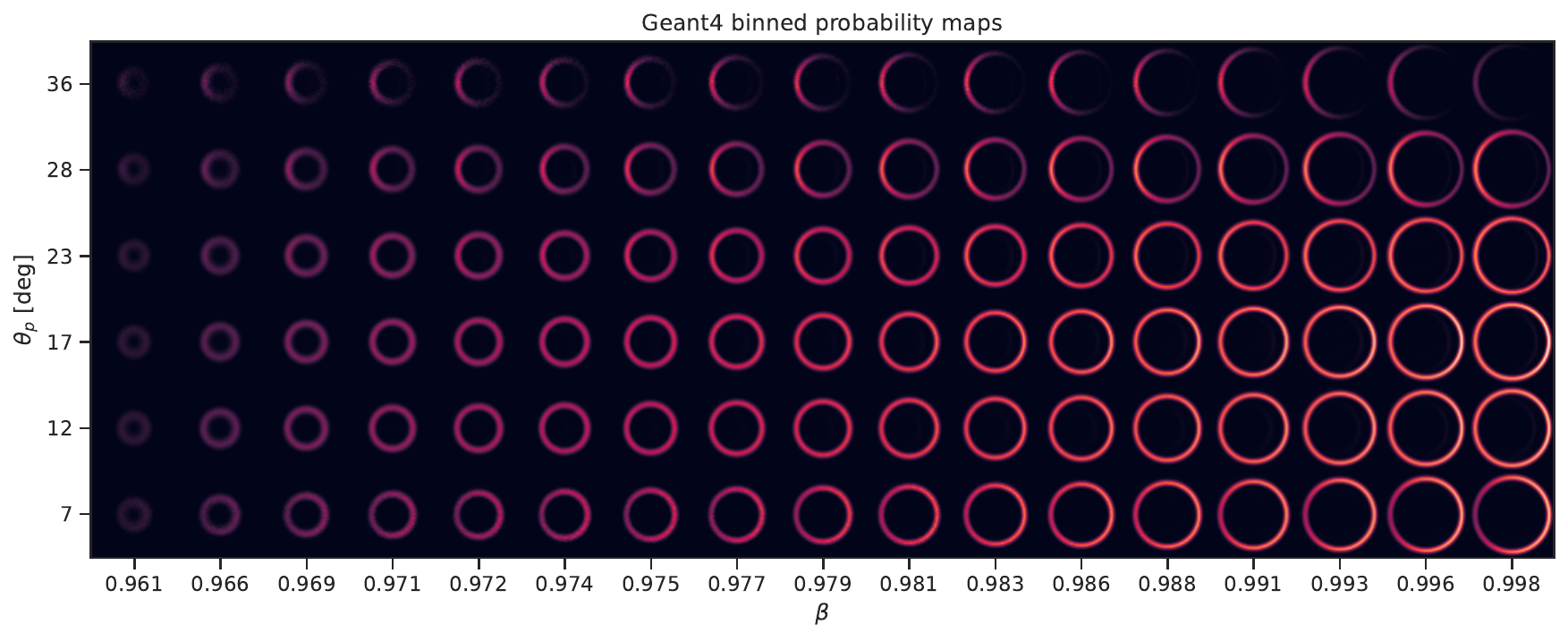}}
\caption{Binned probability maps. Top to bottom: linear baseline, cGAN, Geant4 reference.}
\label{fig:grid_gt}
\alttext{Grid of probability maps.}
\end{figure}

The second metric, $\delta_{\rm RECO}$, is a total variation distance $\delta(P_{\rm Fake}, P_{\rm Geant})$ between
the reconstructed velocity distributions produced by each method and the reference Geant4 simulation. A FARICH reconstruction
algorithm based on $\theta_c$ distribution statistics that was developed earlier\cite{shipilov2024reconstruction} is applied to the outputs of each
simulation method. The resulting $\hat\beta$ distributions are compared
to those obtained from the Geant4 reference.

The cGAN outperforms the
linear baseline in both metrics (see Table~\ref{tab:results}). The MSE for pions and kaons separately shows that kaons are harder to simulate, due to their scarcity in the data and a lower average velocity $\beta$ (higher rest mass). A particularly pronounced improvement is observed
at high $\theta_p$ values, where the linear Gaussian approximation is least accurate (see Figure~\ref{fig:binned}).

Two-dimensional correlation plots of the
reconstructed velocity $\beta_{\rm RECO}$ used in the computation of $\delta_{\rm RECO}$ are demonstrated in the Figure~\ref{fig:corr_pdm}. The $\beta_{\rm RECO}$ spread $\sigma_\beta$ is defined as a standard deviation of $\beta_{\rm RECO} - \beta_{\rm true}$. Notably, the linear model underestimates  $\sigma_\beta$, indicating that it fails to produce hard events that lead to the breakdown of the reconstruction algorithm. cGAN $\sigma_\beta$ is much closer to the correct Geant4 value. The low $\beta$ region also highlights the advantage of cGAN over the baseline.

The cGAN reproduces the ring structure and its variation across bins more faithfully than
the linear model, which is, by construction, limited to rotationally symmetric rings, as demonstrated by the representative grids of
probability maps for a range of $(\beta, \theta_p)$ bins (see Figure~\ref{fig:grid_gt}). Because of this, while the deviation from Geant4 rises for both models as $\theta_p$ rises (see Figure~\ref{fig:binned}), the linear model is more strongly affected. A more sophisticated non-linear baseline may be devised to account for the varying distribution of photon hits along the ring; the authors leave this for future work.


\section{Conclusion}
\label{sec:conclusion}

We present an ML-based fast simulation of the FARICH detector response
using a conditional GAN. The cGAN produces realistic samples and outperforms the linear statistical baseline
by MSE on the probability maps
and total variational distance on the reconstructed $\hat\beta$ distributions. The generator network is lightweight enough to allow a significant throughput gains compared to
Monte-Carlo simulation. Simulating 1\,000\,000 FARICH responses using cGAN takes $~2$ minutes on consumer-grade hardware with further optimizations possible.

The main directions for future work are outlined below. Although the cGAN generation agrees well with Geant4, the reconstructed velocity spread isn't matched perfectly. Further work is needed improve the resolution, possibly by adding a frozen auxiliary regressor to the training procedure to improve reproduction fidelity and conformity to the condition variables. A single step
cGAN generation without intermediate probability maps can be achieved by 
changing the balance between supervised and adversarial components of the loss function,
as well as rewriting the supervised component to penalize aggregate scores instead of individual samples.

\section*{Acknowledgments}

The presented scientific results have been prepared following the materials
of Mirror Laboratories project of HSE University and Partner Organization.

\bibliographystyle{ws-mpla}
\bibliography{Bibliography}

\end{document}